\documentclass[aps,preprint,prb,showpacs,floatfix,superscriptaddress]{revtex4}

\usepackage{epsfig,amsmath,amssymb}
\usepackage{graphics} 
\usepackage{subfig}

\usepackage{multirow}

\begin{document}

\title
{
Magnetic order in a spin-1/2 interpolating kagome-square Heisenberg antiferromagnet
}
\author
{R.~F.~Bishop}
\affiliation
{School of Physics and Astronomy, Schuster Building, The University of Manchester, Manchester, M13 9PL, UK}

\author
{P.~H.~Y.~Li}
\affiliation
{School of Physics and Astronomy, Schuster Building, The University of Manchester, Manchester, M13 9PL, UK}

\author
{D.~J.~J.~Farnell}
\affiliation 
{Health Methodology Research Group, School of Community-Based Medicine, Jean McFarlane Building, University Place, The University of Manchester, M13 9PL, UK}

\author
{C.~E.~Campbell}
\affiliation
{School of Physics and Astronomy, University of Minnesota, 116 Church Street SE, Minneapolis, Minnesota 55455, USA}

\begin{abstract}
  The coupled cluster method is applied to a spin-half model at zero
  temperature ($T=0$), which interpolates between Heisenberg
  antiferromagnets (HAF's) on a kagome and a square lattice. With
  respect to an underlying triangular lattice the strengths of the
  Heisenberg bonds joining the nearest-neighbor (NN) kagome sites are
  $J_{1} \geq 0$ along two of the equivalent directions and $J_{2}
  \geq 0$ along the third. Sites connected by $J_{2}$ bonds are
  themselves connected to the missing NN non-kagome sites of the
  triangular lattice by bonds of strength $J_{1}' \geq 0$. When
  $J_{1}'=J_{1}$ and $J_{2}=0$ the model reduces to the square-lattice
  HAF. The magnetic ordering of the system is investigated and its
  $T=0$ phase diagram discussed. Results for the kagome HAF limit are
  among the best available.
\end{abstract}
\pacs{75.10.Jm, 75.30.Gw, 75.40.-s, 75.50.Ee}
\maketitle

\section{Introduction}
Quantum magnets defined on two-dimensional (2D) spin lattices exhibit
a wide range of physical states at zero temperature, from those with
classical-type ordering (albeit reduced by quantum fluctuations) to
valence-bond solids and spin
liquids.\cite{2D_magnetism_1,2D_magnetism_2} The behavior of these
strongly correlated and often highly frustrated systems is driven by
the nature of the underlying crystallographic lattice, by the number
and range of the magnetic bonds, and by the spin quantum numbers of the
atoms localized to the lattice sites. Very few exact results exist for
such 2D systems and the application of approximate methods has become
crucial to their understanding. A complete picture of their behavior
has only slowly begun to emerge by considering a wide range of
possible scenarios in related models that are themselves often
inspired, or followed closely afterwards, by their experimental
realisation and study. For easy and accurate comparisons to be made it
is clearly preferable to use the same theoretical technique. Among the
most accurate, most universally applicable, and most widely applied to
quantum magnets of such methods is the coupled cluster method
(CCM).\cite{Bi:1991,Ze:1998,Fa:2004} Our aim here is to use the CCM
to extend our understanding of frustrated quantum magnets by applying
it to a novel 2D system that, in some well-defined sense described
below, interpolates between Heisenberg antiferromagnets (HAF's)
defined on square and kagome lattices respectively.

An archetypal and much studied model in quantum magnetism is the
frustrated spin-half $J_{1}$--$J_{2}$ HAF model on the square lattice
with nearest-neighbor (NN) bonds (of strength $J_{1} > 0$) competing
with next-nearest-neighbor (NNN) bonds (of strength $J_{2} \equiv
\alpha J_{1} > 0$). It exhibits two different quasiclassical phases
with collinear magnetic long-range order (LRO) at small ($\alpha <
\alpha_{c_{1}} \approx 0.4$) and large ($\alpha > \alpha_{c_{2}}
\approx 0.6$) values of $\alpha$, separated by an intermediate quantum
paramagnetic phase with no magnetic LRO in the regime $\alpha_{c_{1}}
< \alpha < \alpha_{c_{2}}$. Interest in this model has been
reinvigorated of late by its experimental realisation in such layered
magnetic materials as Li$_2$VOSiO$_4$,\cite{Mel:2000,Ro:2002}
Li$_2$VOGeO$_4$,\cite{Mel:2000} VOMoO$_4$,\cite{Bom:2005} and
BaCdVO(PO$_4$)$_2$.\cite{Na:2008} The syntheses of such layered
quasi-2D materials has stimulated a great deal of renewed interest in
the model (and see, e.g.,
(Refs.~[\onlinecite{Ca:2000,Ca:2001,Sin:2003,Ro:2004}]). Amongst
several methods applied to the $J_{1}$--$J_{2}$ model has been the
CCM.\cite{j1j2_square_ccm1,Bi:2008_PRB,Bi:2008_JPCM,j1j2_square_ccm5,j1j2_square_ccm4}

In view of the huge interest in the $J_{1}$--$J_{2}$ model there have
been several recent attempts to investigate various generalizations and
modifications of the model, in order to shed further light on its
properties. As an example of a generalization of the model we mention
a recent study~\cite{Schm:2006} of the effects of interlayer
couplings on the 2D $J_{1}$--$J_{2}$ model. This study also employed
the CCM in its analysis. Modifications of the $J_{1}$--$J_{2}$ model
that have been studied include models wherein some of the NNN $J_{2}$
bonds are removed. Various such models exist in which either half or
three-quarters of the $J_{2}$ bonds are removed in particular
arrangements, as discussed below. All of these models studied to date
have fascinating magnetic properties and ground-state phases in their
own right.

One such model is the spin-half anisotropic HAF on the 2D
triangular lattice, which has also been studied by the
CCM,\cite{square_triangle} and which interpolates between HAF's on
square and triangular lattices. It is fully equivalent to a variant of
the square-lattice $J_{1}$--$J_{2}$ model in which half of the
$J_{2}$ bonds are removed, leaving just one NNN bond across the same
diagonal of each basic square plaquette. Thus, for this model, the two
cases $J_{2}=0$ and $J_{2} \rightarrow \infty$ relate to a HAF on the
square lattice and a set of decoupled one-dimensional HAF chains
respectively, with the HAF on the triangular lattice in between at
$J_{2}=J_{1}$. Strong evidence was found\cite{square_triangle} that
quantum fluctuations for this spin-half model favor a weakly
first-order (or possibly second-order) transition from N\'{e}el order
to a helical state at a first critical point at $\alpha_{c_{1}} = 0.80
\pm 0.01$ by contrast with the corresponding second-order transition
between the equivalent classical states at $\alpha_{{\rm cl}}=0.5$. The CCM
was also, uniquely, powerful enough to provide strong evidence for a
second quantum critical point at $\alpha_{c_{2}} = 1.8 \pm 0.4$ where
a first-order transition occurs between the helical phase and a
collinear stripe-ordered phase with no classical counterpart, thereby
providing quantitative verification of an earlier qualitative
prediction of such a transition from a renormalization group analysis
of the model.\cite{St:2007} (As a parenthetical note, a different way
of removing half the $J_{2}$ bonds, results in the so-called Union
Jack model, to which the CCM has also recently been
applied,\cite{UJack_ccm} and which has a quite different
zero-temperatue phase diagram).

A further modification of the original spin-half $J_{1}$--$J_{2}$
model is now to remove another half of the $J_{2}$ bonds, leaving half
the fundamental square plaquettes with one $J_{2}$ bond and the other
half with none. One way of doing this in a regular fashion results in
the Shastry-Sutherland model\cite{shastry1} in which no $J_{2}$ bonds
meet at any lattice site and every site is five-connected (by four NN
$J_{1}$ bonds and one $J_{2}$ bond). Interest in this model has been
renewed by the discovery of the magnetic material
$\mathrm{SrCu}_{2}(\mathrm{BO}_3)_2$ [Ref. \onlinecite{Ka:1999}] that can be understood in terms of
it. Its classical ground state is the collinear N\'{e}el state for
$J_{2}/J_{1} \leq 1$ and a noncollinear spiral for $J_{2}/J_{1} > 1$,
with a second-order phase transition in between. However, it is known
that the spin-half model has a quantum ground state which is a product
of local pair singlets (the so-called orthogonal-dimer state) for
$J_{2}/J_{1} \geq 1.465 \pm 0.025$, which has no classical
counterpart. The CCM has also been applied to this
model\cite{shastry2,shastry3} and the latest results\cite{shastry3}
strongly suggest that no intermediate phase exists between the N\'{e}el
and dimerized phases, and that the direct transition between them is
of first-order type.

A different, but equally important, archetypal magnetic system showing
frustration, but now of the geometric kind rather than the dynamic
kind, is the spin-half kagome-lattice HAF. Although this system has
been the subject of intense study over a long period, the nature of
its gound state is still not definitively settled. Among the leading
theoretical contenders are a valence-bond solid
state\cite{Mar:1991,Syr:2002,Nik:2003,Bud:2004,Sin:2007,Eve:2010} and
a spin-liquid
state.\cite{Sach:1992,Leung:1993,Lech:1997,Mila:1998,Wald:1998,Hast:2000,Mamb:2000,Herm:2005,Wang:2006,Ran:2007,Herm:2008,Jiang:2008}
The spin-half kagome-lattice HAF has become the subject of renewed
interest after a possible physical realization of the model has been
found experimentally in the herbertsmithite material
ZnCu$_{3}$(OH)$_{6}$Cl$_{2}$.\cite{Helton:2007,Vdries:2008} A
spatially anisotropic version of the spin-half kagome-lattice HAF has
also been experimentally studied after its physical realization in the
volborthite material
Cu$_{3}$V$_{2}$O$_{7}$(OH)$_{2}$$\cdot$2H$_{2}$O.\cite{Yoshida:2009}
This latter model has also been studied theoretically in recent
years.\cite{Yavo:2007,Wang:2007,Kagome_Schn:2008}

In the present paper we investigate the phase diagram of a novel
spin-half HAF that is another depleted modification of the
$J_{1}$--$J_{2}$ model, and which also contains both the spatially
anisotropic and the isotropic kagome-lattice HAF's discussed above as
limiting cases. As described in more detail below the model also
interpolates continuously between the geometrically frustrated
kagome-lattice HAF and the unfrustrated square-lattice HAF. After
describing the model in Sec. \ref{model_section}, we apply the CCM to
investigate its ground-state properties. The CCM is first described
briefly in Sec. \ref{CCM}, and the results are presented in
Sec. \ref{results}. We conclude in Sec. \ref{discussions_conclusions}
with a discussion of the results and a comparison of them with other
results for limiting cases of our model.

\section{The model}
\label{model_section}
\begin{figure}
\subfloat[Canted\label{fig1a}]{
\begin{minipage}[t]{0.45\linewidth}
\vspace{0pt}
\centering \includegraphics[width=4cm]{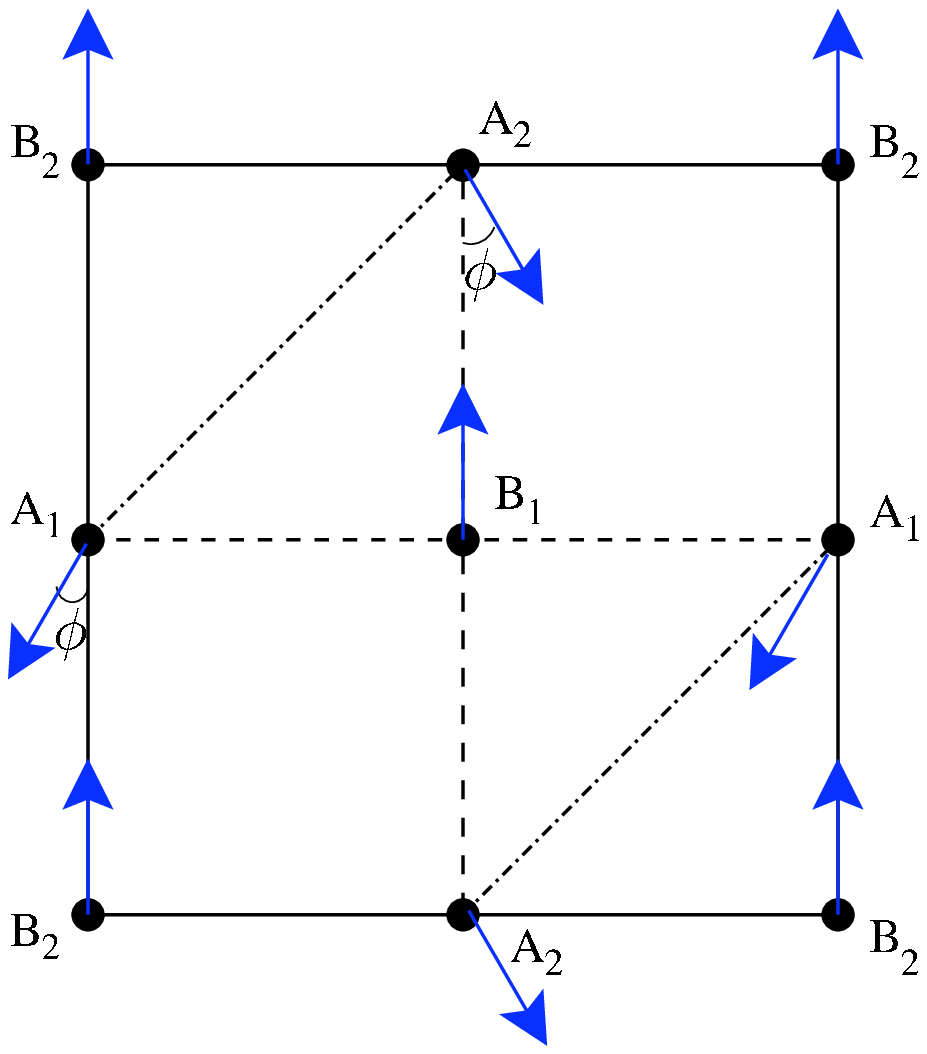}
\end{minipage}
}%
\hfill
\subfloat[Semi-striped\label{fig1b}]{
\begin{minipage}[t]{0.45\linewidth}
\vspace{0pt}
\centering \includegraphics[width=4cm]{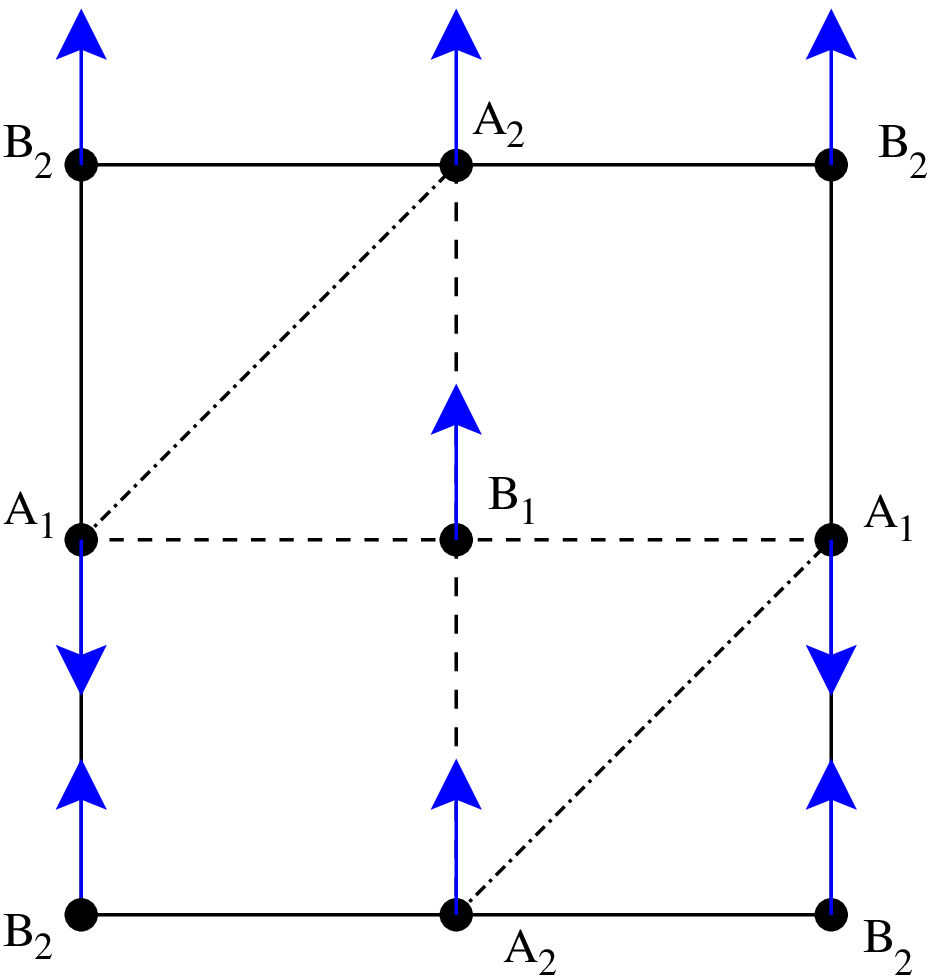}
\vspace{1pt}
\end{minipage}
}%
\caption{(Color online) The interpolating kagome-square model; ---
  $J_{1}$; - - $J_{1}'$; - $\cdot$ - $J_{2}$; showing (a) the canted state; and (b) the semi-striped state.}
\label{fig1}
\end{figure}

In this paper we consider an alternate and novel variant of the
$J_{1}$--$J_{2}$ model in which three-quarters of the $J_{2}$ bonds
are removed from the original $J_{1}$--$J_{2}$ model as in the
Shastry-Sutherland model above, but in a different pattern, as shown
in Fig.~\ref{fig1}.
It may equivalently be obtained from the spin-half anisotropic HAF on
the 2D triangular lattice by removing every second line of NNN $J_{2}$
bonds. The square-lattice representation of the model shown in
Fig. \ref{fig1} contains the two square sublattices of A sites and
B sites respectively, and each of these in turn contains the two
square sublattices of ${\rm A}_{1}$ and ${\rm A}_{2}$ sites, and ${\rm
  B}_{1}$ and ${\rm B}_{2}$ sites respectively, as shown. It is very
illuminating to consider the anisotropic variant in which half of the
$J_{1}$ bonds are allowed to have the strengths $J_{1}' > 0$ along
alternating rows and columns. All of the bonds joining sites $i$ and
$j$ are of standard Heisenberg type, i.e., proportional to {\bf
  s}$_{i}$$\cdot${\bf s}$_{j}$, where the operators {\bf
  s}$_{i}=(s_{i}^{x},s_{i}^{y},s_{i}^{z})$ are the quantum spin
operators on lattice site $i$, with {\bf s}$_{i}^{2}=s(s+1)$ and
$s=\frac{1}{2}$ for the quantum case considered here. 

This so-called {\it interpolating kagome-square model} differs
principally from the Shastry-Sunderland model in that the A sites are
six-connected (by 2 NN $J_{1}$ bonds, 2 NN $J_{1}'$ bonds and 2 NNN
$J_{2}$ bonds) while the B sites are four-connected (by four NN $J_{1}'$
bonds for the B$_{1}$ sites and four NN $J_{1}$ bonds for the B$_{2}$
sites). The spin-half HAF's on the 2D kagome and square lattices are
represented respectively by the limiting cases \{$J_{1}=J_{2},
J_{1}'=0$\} and \{$J_{1}=J_{1}', J_{2}=0$\}. The limiting case when
\{$J_{1}=J_{1}'=0; J_{2}>0$\} represents a set of uncoupled 1D HAF
chains. The case $J_{1}'=0$ with $J_{2} \neq J_{1}$ represents a
spatially anisotropic kagome HAF considered
recently by other authors,\cite{Yavo:2007,Wang:2007,Kagome_Schn:2008} especially in the quasi-1D
limit where $J_{2}/J_{1} \gg 1$.\cite{Kagome_Schn:2008} Henceforth we
set $J_{1} \equiv 1$ and consider the case when all bonds are
antiferromagnetic, (i.e., $J_{1}' \geq 0, J_{2} \geq 0$).

Considered as a classical model (corresponding to the case where the
spin quantum number $s \rightarrow \infty$) the interpolating
kagome-square model has only two ground-state (gs) phases separated by
a continuous (second-order) phase transition at $J_{2}=J_{2}^{{\rm
    cl}} \equiv \frac{1}{2}(J_{1}+J_{1}')$. For $J_{2} < J_{2}^{{\rm
    cl}}$ the system is N\'{e}el-ordered on the square lattice, while
for $J_{2} > J_{2}^{{\rm cl}}$ the system has noncollinear canted
order as shown in Fig.~\ref{fig1}(a), in which the spins on each of
the A$_{1}$ and the A$_{2}$ sites are canted respectively at angles
($\pi \mp \phi$) with respect to those on the B sublattice, all of the
latter of which point in the same direction. The lowest-energy state
in the canted phase is obtained with $\phi = \phi_{{\rm cl}} \equiv
{\rm cos}^{-1}(J_{2}^{{\rm cl}}/J_{2}$). The N\'{e}el state, for
$J_{2} < J_{2}^{{\rm cl}}$, simply corresponds to the case $\phi_{{\rm
    cl}} = 0$. When $J_{1}'=0$ and $J_{2}=J_{1}$, corresponding to the
isotropic kagome-lattice HAF, $\phi_{{\rm cl}}=\frac{1}{3}\pi$, as required by
symmetry. We also note that as $J_{2} \rightarrow \infty$ (with
$J_{1}$ and $J_{1}'$ finite), $\phi_{{\rm cl}} \rightarrow
\frac{1}{2}\pi$, and the spins on the A sublattice become
antiferromagnetically ordered, as is expected, and these spins are
orientated at 90$^{\circ}$ to those on the ferromagnetically-ordered B
sublattice. Of course there is complete degeneracy at this classical
level in this limit between all states for which the relative ordering
directions for spins on the A and B sublattices are arbitrary. The
spin-half problem in the same limit should also comprise decoupled
ferromagnetic and antiferromagnetic sublattices. We expect that this
degeneracy in relative orientation might be lifted by quantum
fluctuations by the well-known phenomenon of {\it
  order-by-disorder}.\cite{Vi:1977} Since it is also true that quantum
fluctuations generally favor collinear ordering, a preferred state is
thus likely to be the so-called ferrimagnetic semi-striped state shown
in Fig.~\ref{fig1}(b) where the A sublattice is now N\'{e}el-ordered
in the same direction as the B sublattice is ferromagnetically
ordered. Alternate rows (and columns) are thus ferromagnetically and
antiferromagnetically ordered in the same direction in the
semi-striped state.

\section{Coupled Cluster Method}
\label{CCM}
The CCM (see, e.g., Refs.~[\onlinecite{Bi:1991,Ze:1998,Fa:2004}] and
references cited therein) that we employ here is one of the most
powerful and most versatile modern techniques available to us in
quantum many-body theory. It has been applied very successfully to
various quantum magnets (see
Refs.~[\onlinecite{j1j2_square_ccm1,Bi:2008_PRB,Bi:2008_JPCM,j1j2_square_ccm4,j1j2_square_ccm5,Fa:2001,Kr:2000,Schm:2006,Fa:2004,Ze:1998,shastry2,shastry3,square_triangle}]
and references cited therein). The method is particularly appropriate
for studying frustrated systems, for which some of the main
alternative methods either cannot be applied or are sometimes only of
limited usefulness, as explained below. For example, quantum Monte
Carlo (QMC) techniques are particularly plagued by the sign problem
for such systems, and the exact diagnoalization (ED) method is
restricted in practice by available computational power, particularly
for $s>1/2$, to such small lattices that it is often insensitive to
the details of any subtle phase order present.

The method of applying the CCM to quantum magnets has been described
in detail elsewhere (see, e.g.,
Refs.~[\onlinecite{Ze:1998,Kr:2000,Fa:2001,Schm:2006,j1j2_square_ccm1,Fa:2004}]
and references cited therein). It relies on building multispin
correlations on top of a chosen gs model state $|\Phi\rangle$ in a
systematic hierarchy of LSUB$n$ approximations (described below) for
the correlation operators $S$ and $\tilde{S}$ that parametrize the
exact gs ket and bra wave functions of the system respectively as
$|\Psi \rangle=e^{S}|\Phi\rangle$ and $\langle \tilde{\Psi}|=\langle
\Phi|\tilde{S}e^{-S}$. In the work presented here we use two different
choices for the model state $|\Phi\rangle$, namely the classical
antiferromagnetic N\'{e}el state and the ferrimagnetic canted
state. We note that the ferrimagnetic semi-striped state provides
another possible choice of model state $|\Phi\rangle$, but we do not
consider it further in the present paper, except in brief remarks
at the end of Sec. \ref{results}.

In each case we employ the well-established LSUB$n$ approximation
scheme in which all possible multi-spin-flip correlations over
different locales on the (square) lattice defined by $n$ or few
contiguous lattice sites are retained. As usual the number of
independent fundamental clusters (i.e., those that are inequivalent
under the symmetries of the Hamiltonian and of the model state)
increases rapidly with the truncation index $n$. For example, the
number of such fundamental clusters for the canted model state is
201481 at the LSUB8 level of approximation in the triangular-lattice
geomery where $J_{2}$ bonds are considered to join NN pairs, and this
is the highest level for the present model that we have been able to attain with available
computing power. In order to solve the corresponding coupled sets of
CCM bra- and ket-state equations we use massively parallel
computing,\cite{ccm} typically using 600 processors simultaneously. We
present results below both at various LSUB$n$ levels of approximation
with $n=\{2,4,6,8\}$ for the case $J_{1}'=J_{1}=1$ and with $n=\{2,4,6\}$
for other values of the bond strengths, and at the corresponding $n
\rightarrow \infty$ extrapolation (LSUB$\infty$) based on the
well-tested extrapolation schemes described below and in more detail
elsewhere.\cite{Ze:1998,Fa:2004,j1j2_square_ccm1,Bi:2008_PRB,Bi:2008_JPCM}
We note that, as always, the CCM exactly obeys the Goldstone
linked-cluster theorem at every LSUB$n$ level of approximation. Hence
we work from the outset in the limit $N \rightarrow \infty$, where $N$
is the number of sites on the square lattice, and extensive quantities
like the gs energy are hence always guaranteed to be linearly
proportional to $N$ in this limit. We note for later purposes that the
number of sites on the kagome lattice (obtained by removing all
B$_{1}$ sites in Fig.~\ref{fig1}) is clearly $N_{K}=\frac{3}{4}N$.
                                                                      
Note that for the canted phase we perform calculations for arbitrary
canting angle $\phi$ shown in Fig.\ \ref{fig1}(a), and then
minimize the corresponding LSUB$n$ approximation for the energy
$E_{{\rm LSUB}n}(\phi)$ with respect to $\phi$ to yield the
corresponding approximation to the quantum canting angle $\phi_{{\rm
    LSUB}n}$. Generally (for $n > 2$) the minimization must be carried
out computationally in an iterative procedure, and for the highest
values of $n$ that we use here the use of supercomputing resources was
essential. Results for the canting angle $\phi_{{\rm LSUB}n}$ will be
given later. 

As always, we choose local spin coordinates on each site for each
choice of model state, so that all spins in $|\Phi\rangle$, whatever
the choice, point in the negative $z$-direction (i.e., downwards) by
definition in these local coordinates.  Then, in the LSUB$n$
approximation all possible multi-spin-flip correlations over different
locales on the lattice defined by $n$ or fewer contiguous lattice
sites are retained. The operator $S$ thus contains only linear sums of
products of creation operator $s^{+}_{k} \equiv s^{x}_{k}+is^{y}_{k}$
on various sites $k$, while the operator $\tilde{S}$ contains only
similar linear sums of products of destruction operators
$s^{-}_{k}\equiv s^{x}_{k}-is^{y}_{k}$. The numbers $N_{f}$ of such
distinct (i.e., under the symmetries of the lattice and the model
state) fundamental configurations of the current model in various
LSUB$n$ approximations are shown in Table~\ref{table_FundConfig}.
\begin{table}
  \caption{Number of fundamental LSUB$n$ configurations ($N_{f}$) for the semi-striped and canted states of the spin-$1/2$ $J_{1}$--$J_{1}'$--$J_{2}$ interpolating square-kagome model.}
\label{table_FundConfig}
\begin{tabular}{ccc} \hline
{Method} & \multicolumn{2}{c}{$N_{f}$} \\ \cline{2-3}
& semi-striped & canted \\ \hline
LSUB$2$ & 3     & 5 \\ 
LSUB$4$ & 32    & 200 \\ 
LSUB$6$ & 645   & 6041 \\ 
LSUB$8$ & 14936 & 201481 \\ \hline
\end{tabular} 
\end{table}
We note that the distinct configurations given in Table
\ref{table_FundConfig} are defined with respect to the geometry
described in Sec.\ \ref{model_section}, and in which the B sublattice
sites of Fig.\ \ref{fig1}(a) are defined to have four NN sites joined
to them by either $J_{1}$ bonds or $J_{1}'$ bonds, and the A sublattice
sites are defined to have the six NN sites joined to them by $J_{1}$,
$J_{1}'$, or $J_{2}$ bonds. If we had chosen instead to work in the
square-lattice geometry every site would have four NN sites.

A significant extra computational burden arises here for the canted
state due to the need to optimize the quantum canting angle $\phi$ at
each LSUB$n$ level of approximation, as described above. Furthermore,
for many model states the quantum number $s^{z}_{T} \equiv
\sum^{N}_{i=1} s^{z}_{i}$ in the original global spin-coordinate
frame, may be used to restrict the numbers of fundamental
multi-spin-flip configurations to those clusters that preserve
$s^{z}_{T}$ as a good quantum number. This is true for the N\'{e}el
state where $s^{z}_{T} = 0$ and for the semi-striped state for which
$s^{z}_{T} = \frac{1}{2}Ns$, where $N$ is the number of lattice
sites. However, for the canted model state that symmetry is absent,
which largely explains the significantly greater number of fundamental
configurations shown in Table \ref{table_FundConfig} for the canted
state at a given LSUB$n$ order. Hence, the maximum LSUB$n$ level that
we can reach here for the canted state, even with massive
parallelization and the use of supercomputing resources, is
LSUB$8$. For example, to obtain a single data point for a given value
of $J_{2}$, with $J_{1}=1$ and $J_{1}'=1$, for the canted phase at the
LSUB$8$ level typically required about 0.2-2.0 h computing time using
600 processors simultaneously for non-critical regions. However, for
values of $J_{2}$ near to critical points, the
LSUB8 computing time increased significantly,
typically to lie in the range of 5-24 h to obtain a single data point
using 600 processors simultaneously.

At each level of approximation we may then calculate a corresponding
estimate of the gs expectation value of any physical observable such
as the energy $E$ and the magnetic order parameter, $M \equiv
-\frac{1}{N}\sum^{N}_{i=1}\langle \tilde{\Psi}|s^{z}_{i}|\Psi
\rangle$, defined in the local, rotated spin axes, and which thus
represents the average on-site magnetization. Note that $M$
is just the usual sublattice (or staggered) magnetization per site for
the case of the N\'{e}el state as the CCM model state, for example.

It is important to note that we never need to perform any finite-size
scaling, since all CCM approximations are automatically performed from
the outset in the infinite-lattice limit, $N \rightarrow \infty$,
where $N$ is the number of lattice sites. However, we do need as a
last step to extrapolate to the exact $n \rightarrow \infty$ limit in
the LSUB$n$ truncation index $n$, at which the complete (infinite) Hilbert space
is reached. We use here the well-tested~\cite{Fa:2001,Kr:2000}
empirical scaling laws
\begin{equation}
E/N=a_{0}+a_{1}n^{-2}+a_{2}n^{-4}\,,  \label{Extrapo_E}
\end{equation} 
\begin{equation}
M=b_{0}+b_{1}n^{-1}+b_{2}n^{-2}\,. \label{Extrapo_M}
\end{equation} 

\section{Results}
\label{results}
\begin{figure}[t]
\epsfig{file=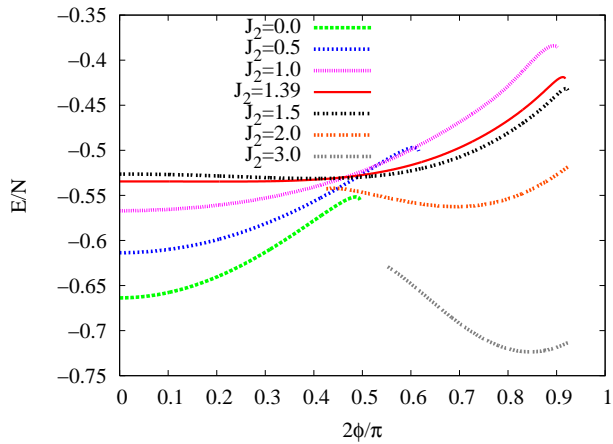,width=6cm,angle=270}
\caption{(Color online) Ground-state energy per spin of the spin-1/2
  interpolating kagome-square model with $J_{1}=J_{1}'=1$, using the
  LSUB4 approximation of the CCM with the canted model state, versus
  the canting angle $\phi$. For $J_{2} \lesssim 1.392$ the minimum is
  at $\phi=0$ (N\'{e}el order) at this level of approximation, whereas
  for $J_{2} \gtrsim 1.392$ the minimum occurs at $\phi=\phi_{{\rm
      LSUB}4} \neq 0$, indicating a phase transition at $J_{2} \approx
  1.392$ in this approximation. Results are shown for those values of
  $\phi$ for which the corresponding CCM equations have real
  solutions.}
\label{fig2}
\end{figure}

We show in Fig.~\ref{fig2}
our CCM results, based on the canted model state, for the gs energy per
spin, $E/N$, plotted as a function of the canting angle $\phi$. We
show results specifically at the LSUB4 level of approximation for the
case $J_{1}=J_{1}'=1$, but our results are qualitatively similar at
other LSUB$n$ levels and for other values of $J_{1}'$ (with
$J_{1}=1$). Curves like those in Fig.~\ref{fig2} show that at this
LSUB4 level of approximation, with $J_{1}=J_{1}'=1$, the minimum
energy is at $\phi = 0$ for $J_{2} < J_{2}^{{\rm LSUB}4} \approx
1.392$ and at a value $\phi \neq 0$ for $J_{2} > J_{2}^{{\rm
    LSUB}4}$. Thus, we have a clear indication of a shift of the
critical point at $J_{2} = J_{2}^{c_{1}}$ between the quantum N\'{e}el
and canted phases from the classical value $J_{2}^{{\rm cl}}=1$ when
$J_{1}=J_{1}'=1$. The observation that N\'{e}el order survives beyond
the classically stable regime, for the quantum spin-half system, is an
example of the promotion of collinear order by quantum fluctuations, a
phenomenon that has been observed in many other systems.

\begin{figure}[t]
\epsfig{file=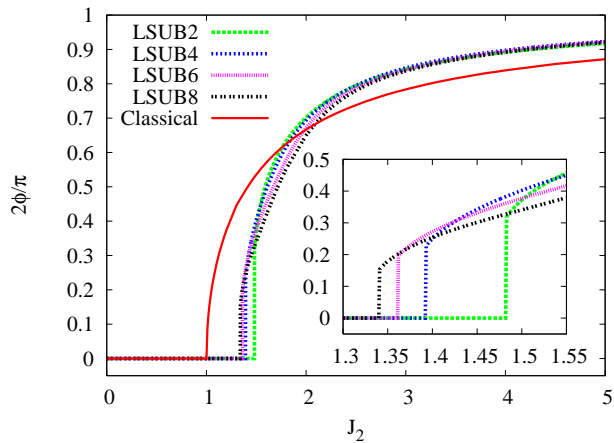,width=6cm,angle=270}
\caption{(Color online) The angle $\phi_{{\rm LSUB}n}$ that minimizes
  the energy $E_{{\rm LSUB}n}(\phi)$ of the spin-half interpolating
  kagome-square model with $J_{1}=J_{1}'=1$, versus $J_{2}$. The
  LSUB$n$ approximations with $n=\{2,4,6,8\}$, using the canted model
  state, are shown. The corresponding classical result $\phi_{{\rm
      cl}}$ is shown for comparison.}
\label{fig3}
\end{figure}

\begin{figure}[t]
\epsfig{file=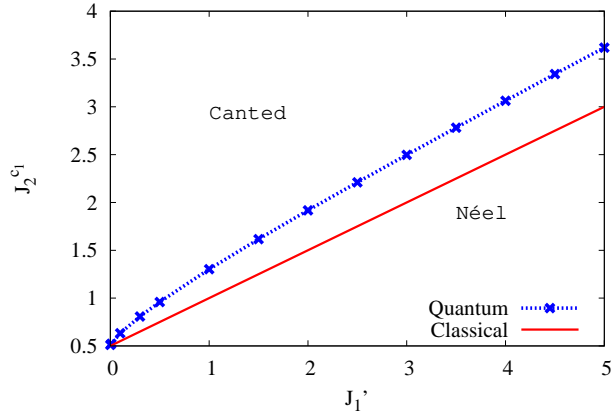,width=6cm,angle=270}
\caption{Ground-state phase diagram of the spin-half interpolating
  square-kagome lattice model, calculated as described in the text,
  compared with its classical counterpart.}
\label{fig4}
\end{figure}

In Fig.~\ref{fig3}
we show the canting angle $\phi_{{\rm LSUB}n}$ that minimizes the gs
energy $E_{{\rm LSUB}n}$($\phi$) at various CCM LSUB$n$ levels based
on the canted state as model state, with $n=\{2,4,6,8\}$, again for
the case $J_{1}'=J_{1}=1$. We see clearly that at each LSUB$n$ level
shown there is a finite jump in $\phi_{{\rm LSUB}n}$ at the
corresponding LSUB$n$ approximation for the phase transition at
$J_{2}=J_{2}^{{\rm LSUB}n}$ between the N\'{e}el state (with
$\phi_{{\rm LSUB}n}=0$) and the canted state (with $\phi_{{\rm LSUB}n}
\neq 0$). Thus at each LSUB$n$ level of approximation the quantum
phase transition is first-order, compared to the second-order
classical counterpart. A close inspection of the insert in
Fig.~\ref{fig3} shows that we cannot completely rule out as $n
\rightarrow \infty$, with increasing level of LSUB$n$ approximation,
the possibility that the phase transition at $J_{2}=J_{2}^{c_{1}}
\equiv J_{2}^{{\rm LSUB}\infty}$ becomes of second-order type,
although a weakly first-order one seems more likely on the evidence so
far. We note that the LSUB$n$ estimates for the phase transition
between N\'{e}el and canted phases, $J_{2}^{{\rm LSUB}n}$, fit well to
an extrapolation scheme $J_{2}^{{\rm LSUB}n} = J_{2}^{{\rm
    LSUB}\infty} + cn^{-1}$. The corresponding estimates for
$J_{2}^{c_{1}}=J_{2}^{{\rm LSUB}\infty}$ in the case shown
($J_{1}=J_{1}'=1$) are $J_{2}^{c_{1}} = 1.298 \pm 0.003$ based on
$n=\{2,4,6,8\}$ and $J_{2}^{c_{1}}=1.302 \pm 0.001$ based on
$n=\{2,4,6\}$ where the errors quoted are those associated with
the specific fit determined via least-squares. We also see from
Fig.~\ref{fig3} that results for the quantum canting angle converge
very rapidly with increasing LSUB$n$ level of approximation, except
for a small region near the phase transition at
$J_{2}=J_{2}^{c_{1}}$. We note too that as $J_{2} \rightarrow \infty$
the canting angle $\phi \rightarrow \frac{1}{2}\pi$ considerably
faster than does the classical analog $\phi_{{\rm cl}}$. Similar
estimates for $J_{2}^{c_{1}}$ have been calculated for other values of
$J_{1}'$ (with $J_{1} \equiv 1$). Thus in Fig.~\ref{fig4}
we compare the phase boundary between the N\'{e}el and canted states
for the present spin-half model with its classical analog. We note in
particular that at $J_{1}'=0$ the critical value calculated as above
is at $J_{2}^{c_{1}} = 0.51 \pm 0.01$, compared with the classical
value $J_{2}^{{\rm cl}} = 0.5$ at $J_{1}'=0$.

\begin{figure}[t]
\epsfig{file=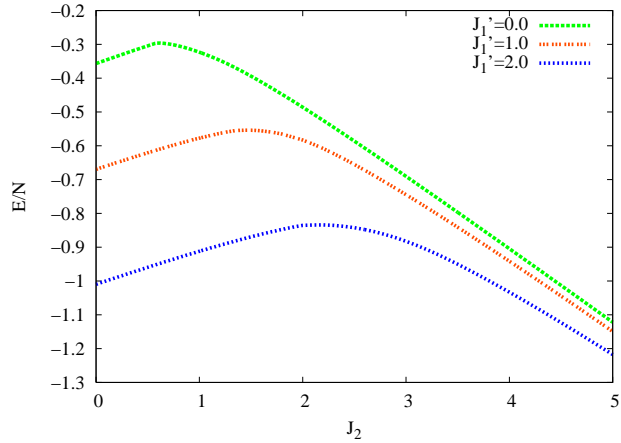,width=6cm,angle=270}
\caption{(Color online) Ground-state energy per spin versus $J_{2}$
  for the N\'{e}el and canted phases of the spin-1/2 interpolating
  kagome-square model ($J_{1}=1$). The CCM LSUB$\infty$ results based
  on LSUB$n$ approximations with $n=\{2,4,6,8\}$ for $J_{1}'=1.0$
  and with $n=\{2,4,6\}$ for the other two $J_{1}'$ values, are shown. The CCM results using the canted model state are based on extrapolated LSUB$n$ approximations with the canting angle $\phi=\phi_{{\rm LSUB}n}$ that minimizes $E_{{\rm LSUB}n}(\phi)$.}
\label{fig5}
\end{figure}

Our CCM results for the gs energy per spin, $E/N$, are shown in
Fig.~\ref{fig5} 
as a function of $J_{2}$, for various values of
$J_{1}'$ (with $J_{1}=1$). (Note that for the case $J_{1}'=0$,
we typically use a very small value $J_{1}'
\approx 10^{-5}$.) At the isotropic kagome point ($J_{1}'=0, J_{2}=J_{1}=1$) the
gs energy per spin is $E/N \approx -0.324 \pm 0.002$, where the error
estimate is based on comparing LSUB$n$ extrapolations from the three
sets $n=\{2,4,6,8\}$, $\{2,4,6\}$, and $\{4,6,8\}$. Expressed
equivalently in terms of the number $N_{K}$ of kagome sites, this
result is $E/N_{K} \approx -0.432 \pm 0.002$. Our corresponding result
for the pure square-lattice HAF (with $J_{1}'=J_{1}=1, J_{2}=0$) is
$E/N \approx -0.6697 \pm 0.0003$. One observes weak signals of a
discontinuity in the first derivative of the energy at the phase
transition points $J_{2}^{c_{1}}$ for all values of $J_{1}'$.

\begin{figure*}[t]
\mbox{
   \subfloat[$J_{1}'=0.25,0.5,1.0,2.0,4.0$]{\scalebox{0.3}{\epsfig{file=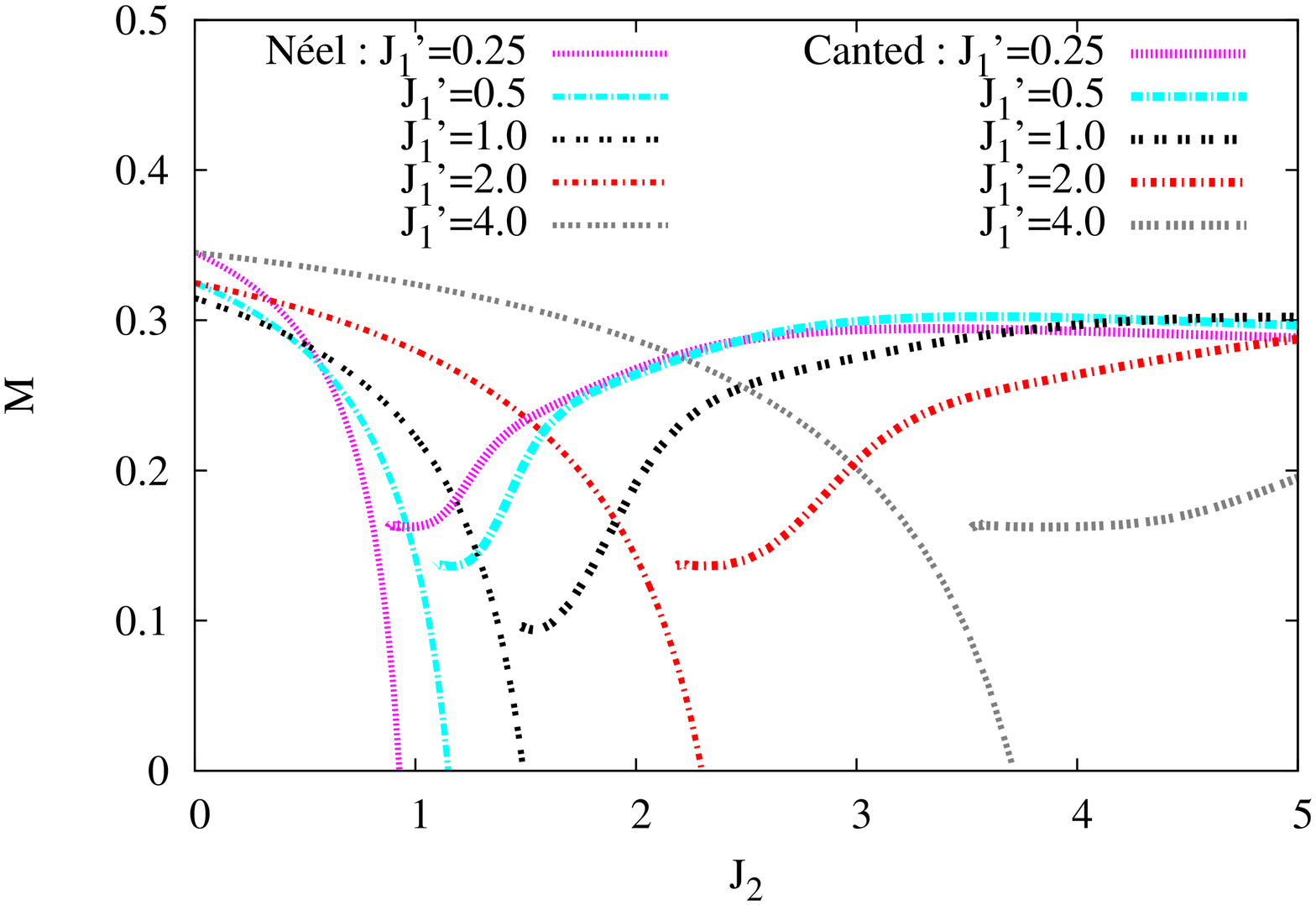,angle=270}}}
}
\mbox{
   \subfloat[$J_{1}'=0.0,0.01,0.1$]{\scalebox{0.3}{\epsfig{file=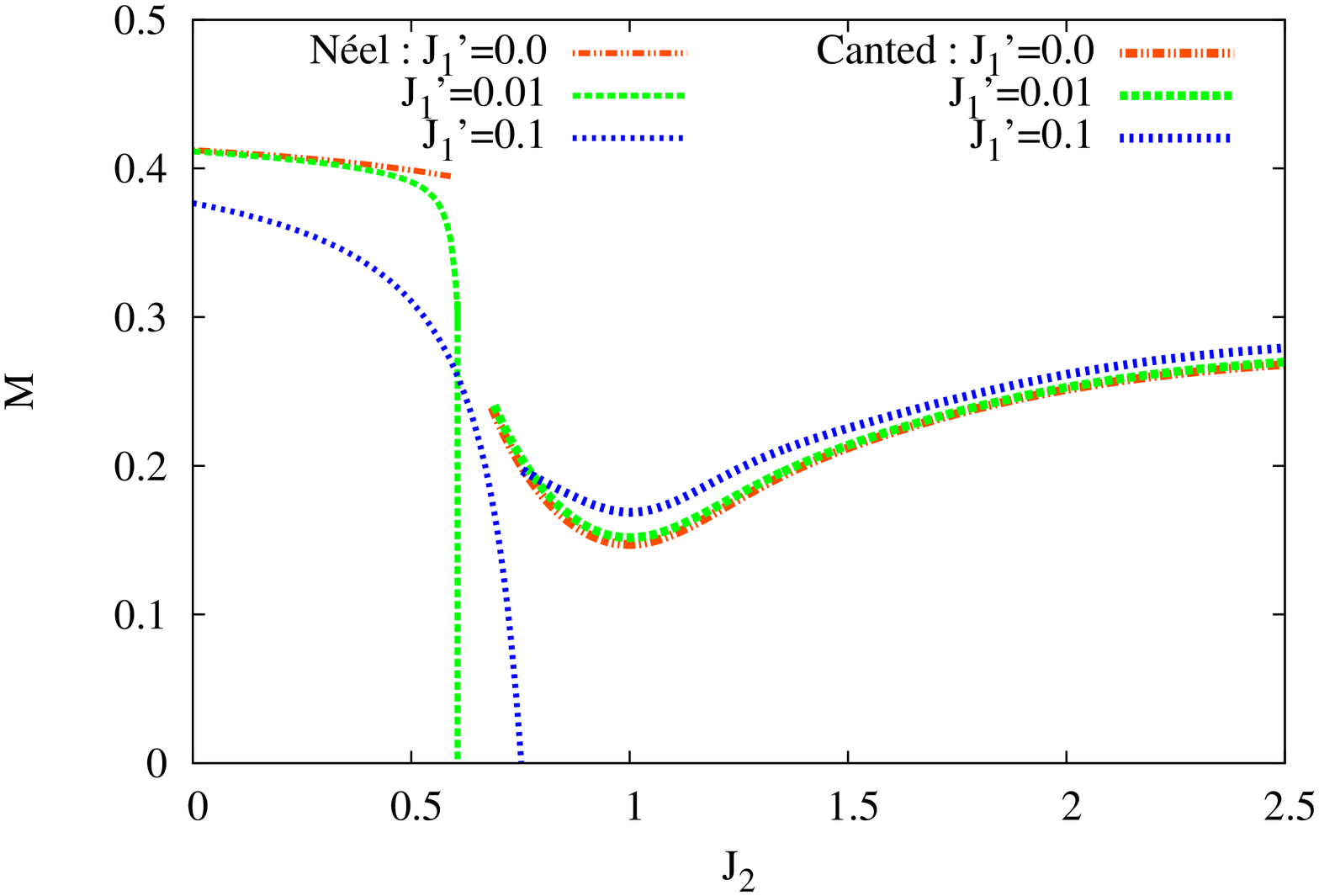,angle=270}}} 
}
\caption{(Color online) Ground-state magnetic order parameter versus
  $J_{2}$ for the N\'{e}el and canted phases of the spin-half
  interpolating kagome-square model ($J_{1}=1$). The CCM LSUB$\infty$
  results based on LSUB$n$ approximations with $n=\{2,4,6,8\}$ for
  $J_{1}'=1.0$ and with $n=\{2,4,6\}$ for the other $J_{1}'$ values,
  are shown. The CCM results using the canted model state are based on
  extrapolated LSUB$n$ approximations with the canting angle
  $\phi=\phi_{{\rm LSUB}n}$ that miniminizes $E_{{\rm LSUB}n}(\phi)$.}
\label{fig6}
\end{figure*}

Much clearer evidence of a first-order phase transition at $J_{2}^{c_{1}}$ is seen in Fig.~\ref{fig6}
where we show comparable CCM results for the average on-site
magnetization or magnetic order parameter, $M$ defined in
Sec. \ref{CCM}. From the symmetry of the model under the interchange
of the bonds $J_{1} \rightleftharpoons J_{1}'$, we note that the order
parameter should satisfy the relation $M(J_{1}, J_{1}', J_{2}) =
M(J_{1}', J_{1}, J_{2}$). Since the order parameter is independent of
an overall scaling of the Hamiltonian, we may also express this
relation in the form $M(1, J_{1}'/J_{1}, J_{2}/J_{1}) = M(1,
J_{1}/J_{1}', J_{2}/J_{1}')$. Thus, placing $J_{1} \equiv 1$, as we
have done here, and considering $M=M(J_{1}'; J_{2})$ as a function of
the remaining two parameters, we have the exact relation $M(J_{1}';
J_{2}) = M(1/J_{1}'; J_{2}/J_{1}'$). The curves shown in
Fig. \ref{fig6}(a) for $J_{1}'=0.25$ and 4.0 and for 0.5 and 2.0 are
easily seen to satisfy this relation for both the N\'{e}el and canted
phases, thereby providing a very good check on our numerics.

For the anisotropic kagome lattice ($J_{1}'=0$), the minimum
LSUB$\infty$ value of $M \approx 0.145$ is seen from
Fig. \ref{fig6}(b) to occur precisely at the isotropic kagome HAF
point ($J_{1}'=0, J_{2}=J_{1}=1$). We may easily re-express this at
the kagome point $J_{1}'=0$ in terms of $M_{K} \equiv
-\frac{1}{N_{K}}\sum^{N_{K}}_{i=1} \langle s^{z}_{i} \rangle$, where
the sum is taken only over the kagome-lattice sites and where again
the spins are defined in the local, rotated spin axes in which all
spins in the CCM model state point in the negative {\it
  z}-direction. Thus, the non-kagome spins on the B$_{1}$ sites are
then frozen in the case $J_{1}'=0$ to have their spins exactly aligned
along the local {\it z}-axes, and hence at the kagome point
$M_{K}=\frac{4}{3}(M-0.125) \approx 0.026$. Our result is thus that
only 5\% of the classical ordering remains for the spin-half kagome
HAF, with an error that makes this compatible with zero. Our
corresponding result for the square-lattice HAF is $M=0.310 \pm
0.003$.

For the N\'{e}el phase ($\phi=0$) curves in Fig. \ref{fig6} we show
the extrapolated (LSUB$\infty$) results for all values of $J_{2}$ (and
given values of $J_{1}$ and $J_{1}'$) for which $M > 0$, and hence
these extend into a regime that is unphysical for the N\'{e}el phase
since the canted phase has lower energy there. By contrast, for the
canted phase ($\phi \neq 0$) curves in Fig. \ref{fig6} we show only
the extrapolated (LSUB$\infty$) results for regimes of $J_{2}$ (and
given values of $J_{1}$ and $J_{1}'$) for which we have LSUB$n$ data
(with $\phi_{{\rm LSUB}n} \neq 0$) for all of the set $n=\{2,4,6\}$,
in order that the $n \rightarrow \infty$ extrapolation can be robustly
performed. Curves like those in Fig. \ref{fig3} thus show that the
results for the canted phase shown in Fig. \ref{fig6} terminate
(artificially) at $J_{2}^{{\rm LSUB}2} > J_{2}^{{\rm LSUB}_{\infty}} =
J_{2}^{c_{1}}$, rather than at the physical value $J_{2}^{c_{1}}$. For
this (unphysical, but computationally imposed) reason, corresponding
pairs of curves for $M$ versus $J_{2}$ (for the same given values of
$J_{1}$ and $J_{1}'$) for the N\'{e}el and canted phases do not
meet. We note, however, that simple (spline) extrapolations of the
LSUB$\infty$ canted-phase magnetization curves generally give 
corresponding estimates for $J_{2}^{c_{1}}$ at which they meet their
N\'{e}el-phase counterparts, which are in excellent agreement with
those calculated as in Fig.~\ref{fig3}. Such extrapolations also give
clear evidence that the order parameter curves for the two phases meet
at a value $J_{2}=J_{2}^{c_{1}}$ at which $M$ is nonzero (and hence
the transition is of first-order type) for all values of $J_{1}'$, and
furthermore that the curves have no discontinuity in slope (or only a
very small one) at $J_{2} = J_{2}^{c_{1}}$. 

We remark that the one region where the extrapolation procedure for
$M$ becomes slightly problematic is for values of $J_{1}' \lesssim
0.1$ near the (generally anisotropic) kagome point, $J_{1}' = 0$. The
reason is clear from Fig.~\ref{fig6}(b) since the N\'{e}el curves drop
to zero from a nonzero value with a slope that approaches infinity as
$J_{1}' \rightarrow 0$. Thus, the N\'{e}el curves for the $J_{1}'=0$
case extend (artificially) to the end-point $J_{2}^{{\rm LSUB}6}
\approx 0.592$, whereas we know the value $J_{2}^{{\rm LSUB}\infty}
\approx 0.51$ in this case. Clearly, the nature of the transition
becomes rather singular in the limit $J_{1}' \rightarrow 0$, as
Fig.~\ref{fig6}(b) clearly shows.

We note finally that the CCM LSUB$n$ solutions with $n > 2$ based on
the canted state terminate at some upper critical value of $J_{2}$ for
all values of $J_{1}'$ (and $J_{1}=1$). This provides preliminary
evidence for another critical point at $J_{2} = J_{2}^{c_{2}}$. For
example, at the isotropic point $J_{1}'=J_{1}=1$ the termination
points occur at values $J_{2} \approx 74.5$, 20.0, and 11.0 for
LSUB$n$ approximations with $n=\{4,6,8\}$ respectively. To investigate
this possible transition further we have performed a preliminary
series of separate CCM LSUB$n$ calculations based on the semi-striped
state shown in Fig.~\ref{fig1}(b). We find that this state is stable
out to the $J_{2} \rightarrow \infty$ limit for all LSUB$n$
approximations investigated ({\it viz.}, with $n \leq 8$). The
corresponding LSUB$\infty$ result for the semi-striped phase as $J_{2}
\rightarrow \infty$ (for fixed $J_{1}'$ and $J_{1}$) is $E/N \approx
-0.2215J_{2}$, which may be compared with the exact result for this
decoupled 1D HAF chain limit (of $N_{c}$ spins per chain) of
$E/(2N_{c}) \approx -0.2216J_{2}$.

We note, however, that the LSUB$\infty$ results for $E/N$ for the
canted and semi-striped phases do not cross for $J_{2} \lesssim 11.0$,
where the canted LSUB$8$ results terminate. Thus, it is not possible
on this evidence alone to suggest that there might be a second
first-order phase transition at $J_{2}=J_{2}^{c_{2}}(J_{1}')$, with
$J_{2} \equiv 1$, between the canted and semi-striped
phases. Nevertheless, the preliminary evidence is that the canted
phase does not exist for values $J_{2} > J_{2}^{c_{2}}(J_{1}')$, with
$J_{1}=1$.

\section{Discussions and Conclusions}
\label{discussions_conclusions}
In this paper we have used the CCM to study the influcence of quantum
fluctuations on the zero-temperature gs properties and phase diagram
of a frustrated spin-half HAF defined on a 2D square lattice with
three sorts of antiferromagnetic Heisenberg bonds of strengths $J_{1}$, $J_{1}'$,
and $J_{2}$ arranged in the pattern shown in Fig. \ref{fig1}. The
$J_{1}$ and $J_{1}'$ bonds are between NN pairs on the square lattice,
while the $J_{2}$ bonds are between only those one-quarter of the NNN
pairs shown. In the case when $J_{1}'=0$ and $J_{2} \neq J_{1}$ the
model reduces to a spin-half HAF on a spatially anisotropic kagome
lattice appropriate to the quasi-2D material volborthite. For the
special case $J_{1}'=0$ and $J_{2}=J_{1}$ the model reduces to the
spin-half HAF on the (isotropic) kagome lattice. Similarly, for the
special case $J_{2}=0$ and $J_{1}'=J_{1}$ the model reduces to the
spin-half HAF on the (isotropic) square lattice. The model thus
interpolates smoothly between these limiting cases.

Classically the model has only two gs phases, namely an
antiferromagnetic N\'{e}el phase and a ferrimagnetic canted phase
shown in Fig. \ref{fig1}(a), and we have focussed attention in the
present work on the effects of quantum fluctuations on these two
classical phases. Consistent with the usual finding that quantum
fluctuations favour collinear configurations of spins, we found that
the phase transition point at $J_{2}^{c_{1}}(J_{1}')$, with $J_{1}
\equiv 1$, between the N\'{e}el and canted phases satisfies the
inequality $J_{2}^{c_{1}}(J_{1}') > J_{2}^{{\rm
    cl}}=\frac{1}{2}(1+J_{1}')$ for all values $J_{1}' \neq 0$, where
$J_{2}^{{\rm cl}}$ is the corresponding classical phase boundary, as
shown in Fig. \ref{fig4}. Precisely at the anisotropic kagome-lattice
point $J_{1}'=0$, however, the classical and quantum critical values
agree to the level of accuracy of our results; $J_{2}^{c_{1}}(0) =
0.51 \pm 0.01$, compared to $J_{2}^{{\rm cl}} = 0.5$.

Our calculations provide strong evidence that the canted phase is not
the stable gs phase for the model for values of $J_{2}$ greater than
some second critical value $J_{2}^{c_{2}}(J_{1}')$ with $J_{1} \equiv
1$. Thus, unlike in the classical case where the canted phase is the
stable gs phase for all values $J_{2} > J_{2}^{{\rm
    cl}}(J_{1}')=\frac{1}{2}(1+J_{1}')$, in the quantum spin-half case
the canted phase seems to be the stable gs phase only for values
$J_{2}^{c_{1}}(J_{1}')<J_{2}<J_{2}^{c_{2}}(J_{1}')$. In order to
investigate the nature of the transition at $J_{2}^{c_{2}}(J_{1}')$
with $J_{1} \equiv 1$, we have also performed preliminary calculations
using the semi-striped state of Fig. \ref{fig1}(b) as model state in
the CCM. Although we have convincing proof that such a semi-striped state is stable
for the spin-half case under quantum fluctuations, for large values of
$J_{2} \rightarrow \infty$ for all values of $J_{1}'$ and $J_{1}
\equiv 1$, its energy is always (very slightly) higher than that of
the canted state in regions where solutions to the corresponding CCM
LSUB$n$ approximations both exist. While these preliminary results do
not exclude a second first-order phase transition at
$J_{2}=J_{2}^{c_{2}}(J_{1}')$, with $J_{1} \equiv 1$, from the canted
phase to the semi-striped phase, it is also quite possible that the
transition at $J_{2} = J_{2}^{c_{2}}(J_{1}')$ is to an entirely
different state. We hope to report further on the existence and nature
of this second quantum phase transition in a future paper.

As stated previously, our main aim here has been to discuss the entire
phase boundary at $J_{2} = J_{2}^{c_{1}}(J_{1}')$, with $J_{1}
\equiv 1$, of the model between the N\'{e}el and canted phases, for
all values of the bond strength $J_{1}'$. Nevertheless, there is no
doubt that the limiting case $J_{1}'=0$, corresponding to the
spatially anisotropic kagome lattice is of huge interest in its own
right, and we aim to discuss this case further in a separate
future paper. The results for the onsite magnetization $M$ shown in
Fig. \ref{fig6} clearly indicate the special nature of the case
$J_{1}'=0$, as we discussed previously in Sec. \ref{results}. Figure
\ref{fig6}(b) shows in particular that the order parameter $M$ at
$J_{1}'=0$ is a minimum for the isotropic kagome lattice
($J_{2}=J_{1}=1$), and we have shown that our results for this
isotropic case are compatible with the vanishing of the corresponding
parameter $M_{K}$ defined on the kagome lattice. Our results for the
gs energy $E/N$ for the isotropic kagome HAF also agree with the best
available by other techniques (and see, e.g., Ref. [\onlinecite{Eve:2010}]).

The isotropic kagome HAF has been greatly studied in the past. The
most direct results from the exact diagonalization of finite
lattices\cite{Lech:1997,Wald:1998} seem to give strong evidence for a
spin-liquid gs phase. Such a conclusion is supported by block-spin
approaches\cite{Mila:1998,Mamb:2000} and by various other
studies.\cite{Sach:1992,Leung:1993,Hast:2000,Herm:2005,Wang:2006,Ran:2007,Herm:2008,Jiang:2008}
Nevertheless, conflicting results have been found by other
authors\cite{Mar:1991,Syr:2002,Nik:2003,Bud:2004,Sin:2007,Eve:2010}
who have proposed various valence-bond solid states as the gs phase of
the isotropic kagome-lattice HAF. A detailed comparison of the exact
spectrum of a 36-site finite lattice sample of the isotropic kagome
HAF against the excitation spectra allowed by the symmetries of
various of the proposed valence-bond crystal states has, however, cast doubts
on their validity.\cite{Misg:2007}

The classical ground states of the anisotropic kagome HAF are spin
configurations that satisfy the condition that for each elementary
triangular plaquette of the kagome lattice in Fig. \ref{fig1} when the
B$_{1}$ sites and $J_{1}'$ bonds are removed (when $J_{1}'=0$), the
energy is minimized. For $J_{2} < \frac{1}{2}$ (with $J_{1}'=0$ and
$J_{1}=1$) the classical ground state is collinear and unique, with
the spins along the $J_{2}$-bond chains aligned in one direction and
the remaining spins on the kagome lattice aligned in the opposite
direction. The total spin of this classical state is thus $S_{{\rm
    tot}}=\frac{1}{3}N_{K}s$ where each spin has magnitude $s$. For
the quantum case the Lieb-Mattis theorem\cite{Lieb:1962} may also be
used, for the limiting case $J_{2}=0$ only, to show that the exact gound
state has the same value $S_{{\rm tot}}=\frac{1}{3}N_{K}s$ of the
total spin as its classical counterpart.

For $J_{2} > \frac{1}{2}$ (with $J_{1}'=0$ and $J_{1}=1$) the
classical ground state is coplanar with the canting angle $\phi={\rm
  cos}^{-1}(\frac{1}{2J_{2}} \neq 0$) shown in Fig. \ref{fig1}. The
classical ensemble of degenerate coplanar states is now characterized
by two variables for each triangular plaquette, namely the angle
$\phi$ such that the middle spin of a given triangular plaquette forms
angles ($\pi \pm \phi$) with the other two spins of the same
plaquette, and the two-valued chirality variable $\chi=\pm 1$ that
defines the direction (anticlockwise or clockwise) in which the spins
turn as one transverses the plaquette in the positive (anticlockwise)
direction. For a given value of $J_{2} > \frac{1}{2}$ (with
$J_{1}'=0$ and $J_{1}=1$) the angle $\phi \neq 0$ is given as above,
and the different degenerate canted states arise from the various
possible ways to assign positive or negative chiralities to the
triangular plaquettes of the lattice.

The HAF on the isotropic kagome lattice (with $J_{1}'=0$ and
$J_{2}=J_{1}=1$) is especially interesting since for this case, with
$\phi=\frac{\pi}{3}$, the number $\Omega$ of degenerate spin
configurations grows exponentially with the number $N_{K}$ of spins,
so that even at zero temperature the system has a nonzero value of the
entropy per spin. By contrast, for the anisotropic case (with
$J_{1}'=0$, $J_{2} \neq J_{1} =1$), the degeneracy $\Omega$ has
been shown\cite{Yavo:2007} to grow exponentially with $\sqrt{N_{K}}$ [i.e., $\Omega \varpropto {\rm exp} (c\sqrt{N_{K}})]$, so that the gs
entropy per spin vanishes in the thermodynamic limit. Clearly, since
in the limit $J_{2} \rightarrow 1$ the anisotropic model approaches
the isotropic model, the anisotropic model must have an appropriately
large number of low-lying excited states that become degenerate with
the ground state in the isotropic limit, $J_{2} \rightarrow 1$.

The spin-half HAF on the spatially anisotropic kagome lattice has been
studied by several authors recently using a variety of
techniques. These have included large-$N$ expansions of the
Sp($N$)-symmetric generalization of the model,\cite{Yavo:2007} a
block-spin perturbation approach to the trimerized kagome
lattice,\cite{Yavo:2007} semiclassical calculations in the limit of
large spin quantum number
$s$,\cite{Yavo:2007,Wang:2007} and field-theoretical
techniques appropriate to quantum critical systems in one dimension
(and which are hence appropriate here for the case $J_{2} \gg J_{1}$
of weakly coupled chains).\cite{Kagome_Schn:2008} The results of such
calculations generally seem to indicate that the anisotropic kagome
HAF (i.e., our model with $J_{1}'=0$) has a N\'{e}el-like gs phase, a
canted coplanar gs phase and, in the limit of large anisotropy ($J_{2}
\gg J_{1} =1$), another gs phase that approaches the decoupled-chain
phase as $J_{2}/J_{1} \rightarrow \infty$. The precise nature of this
third phase is by no means settled, with the results of the various
calculations not in complete agreement with one another. We hope to
contribute our own more detailed CCM results to this debate in the two
future papers outlined above.

\section*{ACKNOWLEDGMENTS}
We thank the University of Minnesota Supercomputing Institute for
Digital Simulation and Advanced Computation for the grant of
supercomputing facilities, on which we relied heavily for the
numerical calculations reported here. We are also grateful for the use
of the high-performance computer service machine (i.e., the Horace
clusters) of the Research Computing Services of the University of
Manchester.


\begin{thebibliography}{99}

\bibitem{2D_magnetism_1} 
U.~Schollw{\"{o}}ck, J.~Richter, D.~J.~J.~Farnell, and R.~F.~Bishop (eds.),
{\em Quantum Magnetism},
Lecture Notes in Physics {\bf 645} (Springer-Verlag, Berlin, 2004).

\bibitem{2D_magnetism_2} 
G.~Misguich and C.~Lhuillier, 
in {\it Frustrated Spin Systems}, edited by H.T.~Diep (World Scientific, Singapore, 2005), p.~229.

\bibitem{Bi:1991} 
R.~F.~Bishop, 
Theor.\ Chim.\ Acta {\bf 80}, 95 (1991).

\bibitem{Ze:1998} 
C.~Zeng, D.~J.~J.~Farnell, and R.~F.~Bishop, 
J.\ Stat.\ Phys. {\bf 90}, 327 (1998).

\bibitem{Fa:2004} 
D.~J.~J.~Farnell and R.~F.~Bishop, 
in {\em Quantum Magnetism}, 
edited by U.~Schollw{\"{o}}ck {\it et al.}, 
Lecture Notes in Physics {\bf 645} (Springer-Verlag, Berlin, 2004), p.307.

\bibitem{Mel:2000} 
R. Melzi, P. Carretta, A. Lascialfari, M. Mambrini, M. Troyer, P. Millet, and F. Mila, 
Phys. Rev. Lett. {\bf 85}, 1318 (2000).

\bibitem{Ro:2002} 
H. Rosner, R.~R.~P. Singh, W.~H. Zheng, J. Oitmaa, S.-L. Drechsler, and W.~E. Pickett, 
Phys. Rev. Lett. {\bf 88}, 186405 (2002).

\bibitem{Bom:2005}
A. Bombardi, L.~C. Chapon, I. Margiolaki, C. Mazzoli, S. Gonthier, F. Duc, and P.~G. Radaelli,
Phys. Rev. B {\bf 71}, 220406(R) (2005).

\bibitem{Na:2008}
R. Nath, A. A. Tsirlin, H. Rosner, and C. Geibel,
Phys. Rev. B {\bf 78}, 064422 (2008).

\bibitem{Ca:2000} 
L. Capriotti and S. Sorella, 
Phys. Rev. Lett. {\bf 84}, 3173 (2000).

\bibitem{Ca:2001} 
L. Capriotti, F. Becca, A. Parola, and S. Sorella,, 
Phys. Rev. Lett. {\bf 87}, 097201 (2001).

\bibitem{Sin:2003}
R.~R.~P. Singh, W. Zheng, J. Oitmaa, O.~P. Sushkov, and C.~J. Hamer, 
Phys. Rev. Lett. {\bf 91}, 017201 (2003).

\bibitem{Ro:2004}
T. Roscilde, A. Feiguin, A.~L. Chernyshev, S. Liu, and S. Haas,
Phys. Rev. Lett. {\bf 93}, 017203 (2004).





\bibitem{j1j2_square_ccm1} 
R.~F.~Bishop, D.~J.~J.~Farnell, and J.~B.~Parkinson, Phys.~Rev.~B {\bf 58},  6394 (1998).


\bibitem{Bi:2008_PRB} 
R.~F.~Bishop, P.~H.~Y.~Li, R.~Darradi, J.~Schulenburg, and J.~Richter,
Phys.\ Rev.\ B {\bf 78}, 054412 (2008).


\bibitem{Bi:2008_JPCM} 
R.~F.~Bishop, P.~H.~Y.~Li, R.~Darradi, and J.~Richter, 
J.\ Phys.: Condens.\ Matter {\bf 20}, 255251 (2008).

\bibitem{j1j2_square_ccm5}
R.~Darradi, O.~Derzhko, R.~Zinke, J.~Schulenburg, S.~E.~Kr\"uger, and J.~Richter, 
Phys.~Rev.~B {\bf 78}, 214415 (2008).

\bibitem{j1j2_square_ccm4}
R.~Darradi, J.~Richter, J.~Schulenburg,
R.~F.~Bishop, and P.~H.~Y.~Li, J.~Phys.: Conf. Ser. {\bf 145}, 012049
  (2009).

\bibitem{Schm:2006} 
D.~Schmalfu$\ss$, R.~Darradi, J.~Richter, J.~Schulenburg, and D.~Ihle, 
Phys.\ Rev.\ Lett. {\bf 97}, 157201 (2006).


\bibitem{square_triangle}
R.~F.~Bishop, P.~H.~Y.~Li, D.~J.~J.~Farnell, and C.~E.~Campbell,
Phys.\ Rev.\ B {\bf 79}, 174405 (2009).

\bibitem{St:2007}
O.~A.~Starykh and L.~Balents,
Phys.\ Rev.\ Lett. {\bf 98}, 077205 (2007).

\bibitem{UJack_ccm}
R.~F.~Bishop, P.~H.~Y.~Li, D.~J.~J.~Farnell, and C.~E.~Campbell,
Phys. Rev. B {\bf 82}, 024416 (2010).

\bibitem{shastry1} 
B.~S.~Shastry and B.~Sutherland, 
Physica B  {\bf 108}, 1069 (1981).

\bibitem{Ka:1999} 
H. Kageyama, K. Yoshimura, R. Stern, N.~V. Mushnikov, K. Onizuka, M. Kato, K. Kosuge,
C.~P. Slichter, T. Goto, and Y. Ueda,
Phys. Rev. Lett. {\bf 82}, 3168 (1999).


\bibitem{shastry2} R. Darradi, J. Richter, and D.~J.~J. Farnell, Phys. Rev. B. {\bf 72}, 104425 (2005).

\bibitem{shastry3} D.~J.~J. Farnell, J. Richter, R. Zinke, and R.~F. Bishop, J.~Stat.~Phys.~{\bf 135}, 175 (2009)



\bibitem{Mar:1991}
J.~B. Marston and C. Zeng, 
J. Appl. Phys. {\bf 69}, 5962 (1991).

\bibitem{Syr:2002}
A.~V. Syromyatnikov and S.~V. Maleyev,
Phys. Rev. B {\bf 66}, 132408 (2002).

\bibitem{Nik:2003}
P. Nikolic and T. Senthil,
Phys. Rev. B {\bf 68}, 214415 (2003).

\bibitem{Bud:2004}
R. Budnik and A. Auerbach,
Phys. Rev. Lett. {\bf 93}, 187205 (2004).

\bibitem{Sin:2007}
R. R. P. Singh and D. A. Huse,
Phys. Rev. B {\bf 76}, 180407(R) (2007); {\it ibid}. {\bf 77}, 144415 (2008).

\bibitem{Eve:2010}
G. Evenbly and G. Vidal,
Phys. Rev. Lett. {\bf 104}, 187203 (2010).

\bibitem{Sach:1992}
S. Sachdev,
Phys. Rev. B {\bf 45}, 12377 (1992).

\bibitem{Leung:1993}
P. W. Leung and V. Elser,
Phys. Rev. B {\bf 47}, 5459 (1993).

\bibitem{Lech:1997}
P. Lecheminant, B. Bernu, C. Lhuillier, L. Pierre, and P. Sindzingre,
Phys. Rev. B {\bf 56}, 2521 (1997).

\bibitem{Mila:1998}
F. Mila,
Phys. Rev. Lett. {\bf 81}, 2356 (1998).

\bibitem{Wald:1998}
C. Waldtmann, H.-U. Everts, B. Bernu, C. Lhuillier, P. Sindzingre, P. Lecheminant, and L. Pierre,
Eur. Phys. J. B {\bf 2}, 501 (1998).

\bibitem{Hast:2000}
M. B. Hastings,
Phys. Rev. B {\bf 63}, 014413 (2000).

\bibitem{Mamb:2000}
M. Mambrini and F. Mila,
Eur. Phys. J. B {\bf 17}, 651 (2000).

\bibitem{Herm:2005}
M. Hermele, T. Senthil, and M. P. A. Fisher,
Phys. Rev. B {\bf 72}, 104404 (2005).


\bibitem{Wang:2006}
F. Wang and A. Vishwanath,
Phys. Rev. B {\bf 74}, 174423 (2006).

\bibitem{Ran:2007}
Y. Ran, M. Hermele, P. A. Lee, and X.-G. Wen,
Phys. Rev. Lett. {\bf 98}, 117205 (2007).

\bibitem{Herm:2008}
M. Hermele, Y. Ran, P. A. Lee, and X.-G. Wen,
Phys. Rev. B {\bf 77}, 224413 (2008).

\bibitem{Jiang:2008}
H. C. Jiang, Z. Y. Weng, and D. N. Sheng,
Phys. Rev. Lett. {\bf 101}, 117203 (2008).

\bibitem{Helton:2007}
J. S. Helton, K. Matan, M. P. Shores, E. A. Nytko, B. M. Bartlett, 
Y. Yoshida, Y. Takano, A. Suslov, Y. Qiu, J.-H. Chung, D. G. Nocera, and Y. S. Lee,
Phys. Rev. Lett. {\bf 98}, 107204 (2007).

\bibitem{Vdries:2008}
M. A. de Vries, K. V. Kamenev, W. A. Kockelmann, J. Sanchez-Benitez, and A. Harrison,
Phys. Rev. Lett. {\bf 100}, 157205 (2008).

\bibitem{Yoshida:2009}
M. Yoshida, M. Takigawa, H. Yoshida, Y. Okamoto, and Z. Hiroi,
Phys. Rev. Lett. {\bf 103}, 077207 (2009).

\bibitem{Yavo:2007}
T. Yavors'kii, W. Apel, and H.-U. Everts,
Phys. Rev. B {\bf 76}, 064430 (2007).

\bibitem{Wang:2007}
F. Wang, A. Vishwanath, and Y. B. Kim,
Phys. Rev. B {\bf 76}, 094421 (2007).

\bibitem{Kagome_Schn:2008}
A.~P.~Schnyder, O.~A.~Starykh, and L.~Balents,
Phys.\ Rev.\ B {\bf 78}, 174420 (2008).

\bibitem{Vi:1977} 
J.~Villain, 
J.\ Phys.\ (France) {\bf 38}, 385 (1977); 
J.~Villain, R.~Bidaux, J.~P.~Carton, and R.~Conte, 
{\it ibid.} {\bf 41}, 1263 (1980). 



\bibitem{Fa:2001}
D.~J.~J.~Farnell, R.~F.~Bishop, and K.~A.~Gernoth,
Phys.\ Rev.\ B {\bf 63}, 220402(R) (2001).

\bibitem{Kr:2000} 
S.~E.~Kr{\"{u}}ger, J.~Richter, J.~Schulenburg, D.~J.~J.~Farnell, and R.~F.~Bishop, 
Phys.\ Rev.\ B {\bf 61}, 14607 (2000).



\bibitem{ccm} We use the program
package CCCM of D.~J.~J. Farnell and J. Schulenburg, see
http://www-e.uni-magdeburg.de/jschulen/ccm/index.html.

\bibitem{Misg:2007}
G. Misguich and P. Sindzingre, 
J. Phys.: Condens. Matter {\bf 19}, 145202 (2007).

\bibitem{Lieb:1962}
E. Lieb and D. Matthis, 
J. Math. Phys. {\bf 3}, 749 (1962).


\end{thebibliography}
\end{document}